\documentclass[manuscript]{aastex}

\newcommand{\arcm}{\mbox{$^{\prime}$}}
\newcommand{\arcs}{\mbox{$^{\prime\prime}$}}

\shorttitle{NIR Galaxy Counts and Color Distributions}
\shortauthors{Imai et al.}

\begin{document}
\title{$J$- and $K_S$-band Galaxy Counts and Color Distributions in the $AKARI$ North Ecliptic Pole Field}

\author{Koji Imai, Hideo Matsuhara, Shinki Oyabu,\\
Takehiko Wada, Toshinobu Takagi, Naofumi Fujishiro}
\affil{Institute of Space and Astronautical Science,
Japan Aerospace Exploration Agency, Sagamihara, Kanagawa 229 8510}
\email{koji@ir.isas.jaxa.jp}

\author{Hitoshi Hanami}
\affil{Iwate University}

\and

\author{Chris P. Pearson}
\affil{European Space Astronomy Centre (ESAC), Apartado 50727, 28080 Madrid, Spain\\
Institute of Space and Astronautical Science,
Japan Aerospace Exploration Agency, Sagamihara, Kanagawa 229 8510}

\begin{abstract}
We present the $J$- and $K_S$-band galaxy counts and galaxy colors
covering 750 square arcminutes in the deep $AKARI$ North Ecliptic Pole (NEP) field,
using the FLoridA Multi-object Imaging Near-ir Grism Observational Spectrometer (FLAMINGOS)
on the Kitt Peak National Observatory (KPNO) 2.1m telescope. 
The limiting magnitudes with a signal-to-noise ratio of three in the deepest regions are
21.85 and 20.15 in the $J$- and $K_S$-bands respectively in the $Vega$ magnitude system.
 
The $J$- and $K_S$-band galaxy counts in the $AKARI$ NEP field are broadly in good agreement with those of other results in the literature,
however we find some indication of a change in the galaxy number count slope at $J\sim19.5$ and over the magnitude range $18.0<K_S<19.5$. 
We interpret this feature as a change in the dominant population at these magnitudes
because we also find an associated change in the $B-K_S$ color distribution at these magnitudes
where the number of blue samples in the magnitude range $18.5<K_S<19.5$ is significantly larger than that of $K_S<17.5$.
\end{abstract}


\section{Introduction}
The infrared universe still holds many unresolved questions: for example, the formation and evolution of galaxies and large scale structure.
To answer such questions, JAXA (Japan Aerospace eXploration Agency) launched an infrared astronomy satellite 
$AKARI$ \citep[formerly ASTRO-F,][]{mur04} on 22 February, 2006 (Japan Standard Time, JST).
$AKARI$ is equipped with a 68.5 cm cooled telescope and two scientific instruments,
namely the Far-Infrared Surveyor \citep[FIS,][]{kaw04} and the Infrared Camera \citep[IRC,][]{ona04}.
A major goal of the mission is to perform an all-sky survey in six infrared wavebands from 9-180 microns
to higher sensitivity, spatial resolution and larger wavelength coverage
than the first such survey made by the Infrared Astronomical Satellite \citep[IRAS,][]{hac87}.
In addition $AKARI$ will perform deep surveys over the wavelength range of 1.7 microns (near-infrared) to 180 microns (far-infrared).
Since the orbit of $AKARI$ is Sun-synchronous polar, $AKARI$ can often obtain deep exposures at the ecliptic poles.
Such a deep survey program currently underway in the $AKARI$ North Ecliptic Pole (NEP) field \citep{mat06}.
Our near-infrared (NIR) ground survey is designed to identify the counterparts of sources detected with the $AKARI$ NEP deep survey.
We will study the nature of the detected sources via examination of their broadband energy distributions by also using the optical deep Subaru images in the same field (Wada et al. 2007 in preparation).
In this paper, we mainly discuss the $J$- and $K_S$-band galaxy counts and compare these with the results from other fields.

The galaxy counts as a function of magnitude are a basic quantity in the investigation of galaxy evolution.
It has been studied extensively in the optical region.
However the NIR galaxy counts provide more accurate evolutional information on galaxies,
since the effects of dust extinction and star formation are less severe than in the optical, hence the NIR light traces the underlying stellar mass more closely.
Furthermore $K$-corrections due to the redshifted spectral energy distributions are also smaller and nearly independent of galaxy type \citep{cow96, pog97}.

For over a decade, there have been numerous surveys in the NIR $K$-band.
The results of these galaxy counts have stimulated a lot of debates.
The first claim against the single power-low slope of the $K$-band galaxy counts was reported by \citet{gar93}.
From four blank-field surveys (HWS, HMWS, HMDS, HDS), 
they showed that the slope of their counts changed at $K\sim17$ from $d(\log{N})/dm\sim0.67$ to 0.26.
In contrast, subsequent deeper counts with the Keck 10 m telescope \citep{djo95} did not show any evidence for a turnover down to a limiting magnitude of $K\sim24$.
However a superposition of the available results for wide and deep counts indeed showed a bump in the count slope at $16<K<20$ \citep{sar97}.
\citet{cri03} point out that reproducing the change in the slope at $K_S\sim17.5$ requires a delay in the galaxy formation epoch for both elliptical and 
spiral galaxies to lower values, $z_{form}<2$ and the presence of a population of star-forming dwarfs at all redshifts in a $\Lambda$-dominated universe.
The tendency of the bump is also seen in recent results from wide and deep surveys,
though the fainter counts of wide field surveys \citep[e.g. 7.1 deg$^2$][]{els06} require a correction for completeness
while the brighter part of counts of deep surveys \citep[e.g.][]{roc03,cap05,for06} have large errors due to the lower number of galaxies.
It was also reported that galaxy populations below the turning point of the $K$-band count slope tend to have bluer optical-NIR colors \citep[e.g.][]{gar93,gar95,djo95,sar97,ber98,sar99,mcc00,hua01,kum01}.
On the other hand, papers describing the $J$-band galaxy counts published to date number only a few. 
The deep counts with the Keck 10 m telescope \citep{ber98} and wide field surveys covering a total area of roughly 1 deg$^2$ \citep{vai00,dro01}
did not show any sign of a flattening in the slope down to their limiting magnitudes.
It is important to examine the existence any such change of galaxy counts and corresponding color by using the data obtained from improved deep, wide-field imaging surveys,
since these will be fundamental clues to understand the evolution in the galaxy populations.

The $K_S$-band survey in the $AKARI$ NEP field covers the magnitude range $16.5<K_S<19.5$, so is therefore useful for determining the exact location of the change in galaxy count slope and the change of galaxy color.
On the other hand, the $J$-band survey covers the magnitude range $17.0<J<20.5$, and is therefore valuable to link other deep and wide galaxy counts.
We also make an attempt to discover any corresponding change or confirm the unreported change in the $J$-band galaxy count slope by using our larger galaxy sample. 

We describe the observations and reduction strategy in \S \ref{sec:observations&data}.
We then present our procedure to separate stars and galaxies,
completeness corrections and completeness corrected galaxy counts in \S \ref{sec:galaxy_counts}.
The galaxy color trend and the color distribution are describe in \S \ref{sec:galaxy_color}.
In $\S \ref{sec:discussion}$, we discuss the source of the uncertainties in the galaxy counts,
possible causes of observed change in the galaxy counts and the bluer trend in the $B-K_S$ color distribution.
Finally, the summary of this research is given in \S \ref{sec:sammary}. 
Throughout this paper a cosmology with H$_0=70$ kms$^{-1}$Mpc$^{-1}$,
$\Omega_{m}=0.3$ and $\Omega_{\Lambda}=0.7$ is assumed in agreement with the results for Type Ia supernovae \citep{rie98,per99} and WMAP \citep{spe03}.
Unless otherwise noted, we use the $Vega$ magnitude system.

\section{Observations and Data}
\label{sec:observations&data}
\subsection{Observations}
\label{sub:observations}
\begin{figure}
\epsscale{.60}
\plotone{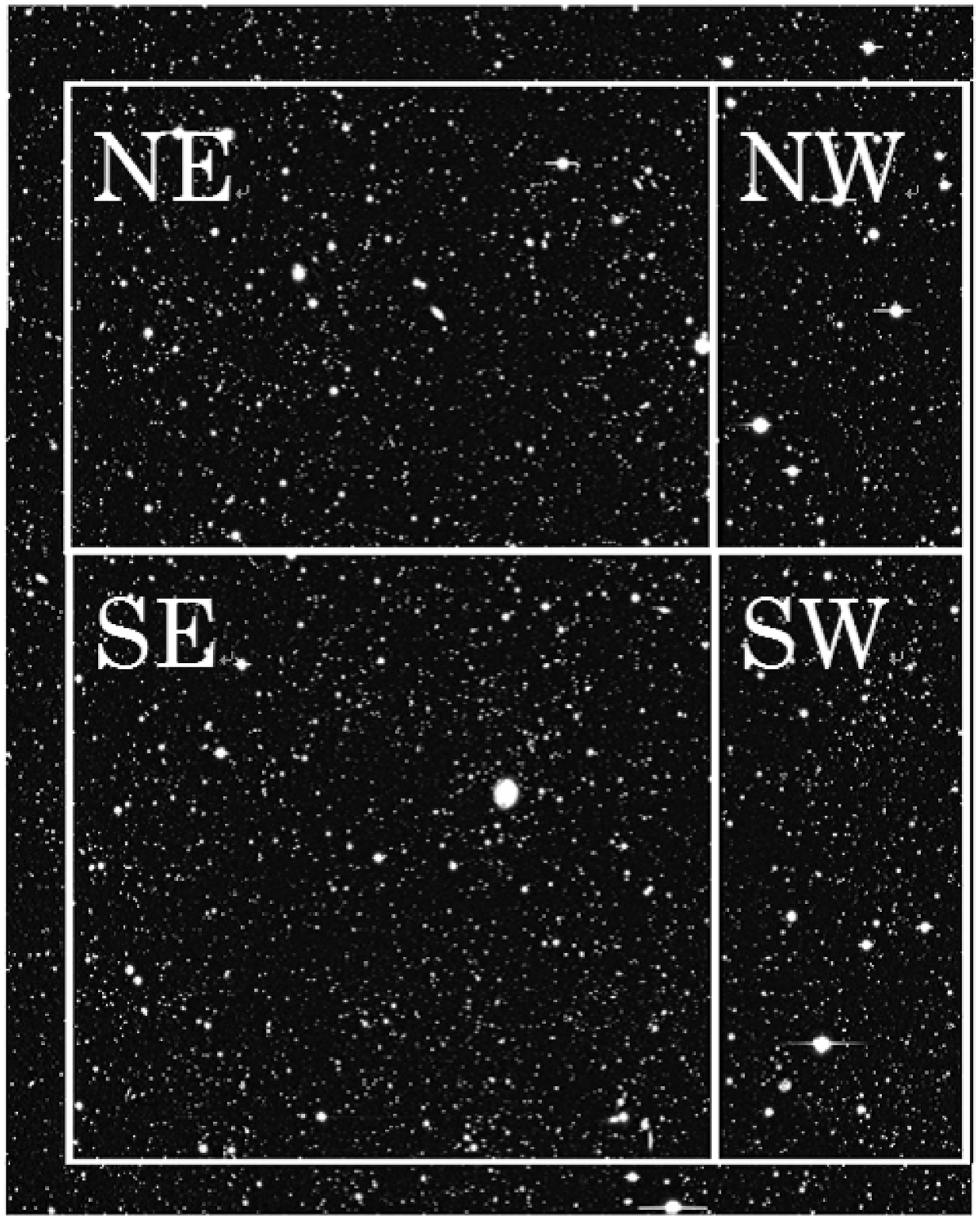}
\caption{Final coverage for our NIR survey in the $AKARI$ NEP field.
The four segments marked by solid square boxes with white lines are overlaid on the $z'$-band image (Wada et al. 2007 in preparation). 
The top-left, top-right, bottom-left and bottom-right square boxes are referred to as NE, NW, SE and SW, respectively.
The center coordinate and area of these observations are listed in Table \ref{table:obs_area}.
\label{figure:SCAMvsFLMN_at_NEP}}
\end{figure}

The $J$- and $K_S$-band imaging were made with the FLAMINGOS \citep{els98,els03} on the KPNO 2.1 m telescope in 2004 June 12-16.
The detector is a Hawaii II 2048$\times$2048 HgCdTe science grade array divided into four quadrants with eight amplifiers each.
The FOV is 20.7\arcm\ with pixel scale of 0.606\arcs.\
To cover the entire field of the optical Suprime-CAM images
(of which details are given in Wada et al. 2007 in preparation,
four different pointings, which are named as NE, NW, SE and SW, were used in each band.
The total effective area is 750 arcmin$^2$ in both the $J$- and $K_S$-bands.
Figure \ref{figure:SCAMvsFLMN_at_NEP} show these divided survey areas on the $z'$-band image of the $AKARI$ NEP field.
The center coordinate and area of these observations are listed in Table \ref{table:obs_area}.

The observation in each band and position follows the standard
sequence which consists of the 5$\times$5 dither pattern with 20\arcs offsets.
A single exposure time of 120s was employed throughout the observations in the $J$-band,
while the 20s and 30s exposure times in the $K_S$-band were used to adjust to the background
caused by the thermal emission from warm optics and telescope structure.
The total exposure time in the $J$-band for each segment was 9480s, 8400s, 13080s, and 10920s in the NE, NW, SE and SW, respectively
and in the $K_S$-band was 6160s, 7110s, 6950s and 7980s in the NE, NW, SE and SW, respectively.
The sky condition during this observation was photometric and we also confirmed that the fluxes of sources on the data were stable through the observation.
The stellar image size of these regions ranged from 1.08\arcs\ to 1.86\arcs\ in FWHM for the $J$-band,
and from 1.08\arcs\ to 2.04\arcs\ in FWHM for the $K_S$-band.
The standard stars, 9169 and 9177 listed in \citet{per98} were observed for flux calibration.
Uncertainties in the flux calibration are estimated to be less than $\pm 0.1$mag.
Table \ref{table:obs_result} summarizes the observational conditions i.e., the total exposure time,
FWHM of point sources, the 50\% completeness limit, and the 3$\sigma$ limiting magnitude for each position. 

\subsection{Reduction}
\label{sub:reduction}
Data reduction was based on the IRAF data reduction package
\footnote{IRAF is distributed by the National Optical Astronomy Observatories,
which are operated by the Association of Universities for Research in Astronomy, Inc.,
under cooperative agreement with the National Science Foundation.}.
Dark-averaged images for each exposure time were made and subtracted from all images.
The sky flat images made with a lot of dark sky images were not practical
because they were not free from any thermal component coming from the warm optics, and telescope structure.
Thus flat-field calibration images were made by the subtraction of lamp-off dome flat images from lamp-illuminated dome flat images.
These flat-field images are used to flatten all the individual images.

A external IRAF package, the Experimental Deep Infrared Mosaicing Software \citep[XDIMSUM,][]{sta95}
are used to perform careful sky subtraction, bad-pixel and cosmic ray masking,
and final combination of the sky-subtracted individual images.
We describe the details below.

In the first stage,
the `median' sky image made with five neighboring images were used to subtract for the sky component of each image
which also consists of the thermal component from the warm optics and telescope structure.
The possible cosmic ray events or rare pixel events due to unstable behavior and tracks of artificial satellites
were excluded by rejecting the pixel values with scatters exceeding 2$\sigma$ over the entire images.
Then, we generated the reduced images in each band and
each field in the sky,
stacking all of the sky-subtracted frames with relative offsets which
had been determined from the location of bright sources.
From the first reduced image, we performed source extraction
to make a object catalog as the input of the second stage reduction.
This catalog was used to make the object masks for each image.
Then we repeated the same steps of the first stage;
sky subtraction, bad-pixel and cosmic rays masking, and final stacking of the sky-subtracted individual images.
The difference from the first stage was to make much more accurate sky
images for the sky subtraction.
To make the new `median' sky image of five neighboring images,
We masked out the objects some of which could not be recognized on each image.
Thus we succeeded with an accurate sky subtraction using the new sky images which were free from faint objects.
Final stacked images of the second stage were the final products
in this data reduction.

For source detection and photometry on the reduced images,
We employed the SExtractor routine, developed by \citet{ber96}.
Detection was calculated using the rms map generated during the image combination process.
Objects were detected in the condition with four pixels rising more than 1.5$\sigma$ above the sky background rms noise,
but only those brighter than 3$\sigma$ limiting magnitude in Table \ref{table:obs_result}
were listed in the source catalog.

\section{$J$- and $K_S$-Band Galaxy Counts}
\label{sec:galaxy_counts}
\subsection{Star-galaxy separation}
\label{sub:star-galaxy_separation}
Star-galaxy separation is important especially at bright magnitudes,
since the modeling of the faintest galaxy counts to study the evolution of distant galaxy populations 
relies heavily on an accurate normalization of the models at brighter magnitudes.
However the spatial resolution of the NIR images is not good enough for a morphological separation of stars and galaxies to be secure.
Instead, the detected objects were classified on the basis of the image profile of their optical counterparts (Wada et al. 2007 in preparation).

\begin{figure}
\epsscale{.60}
\plotone{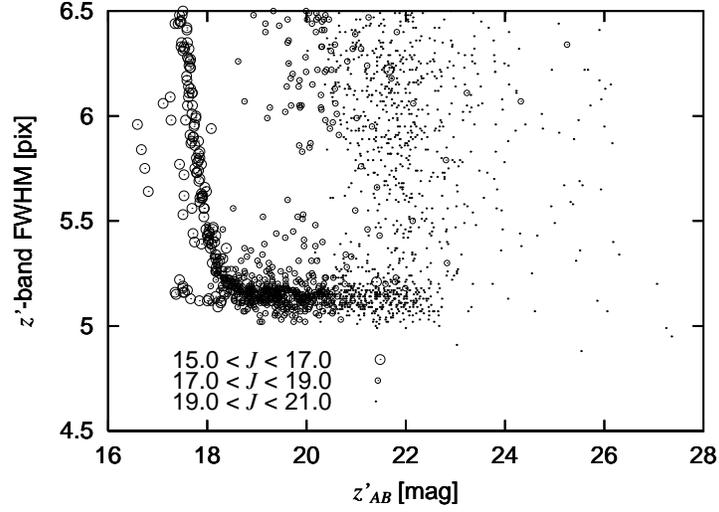}
\caption{$z'$-band FWHM vs. magnitude with the corresponding $J$-band objects.
The point-like objects at the seeing limit cluster together around a FWHM of 5.2 pixels.
\label{figure:saturation_J}}
\end{figure}

\begin{figure}
\epsscale{.60}
\plotone{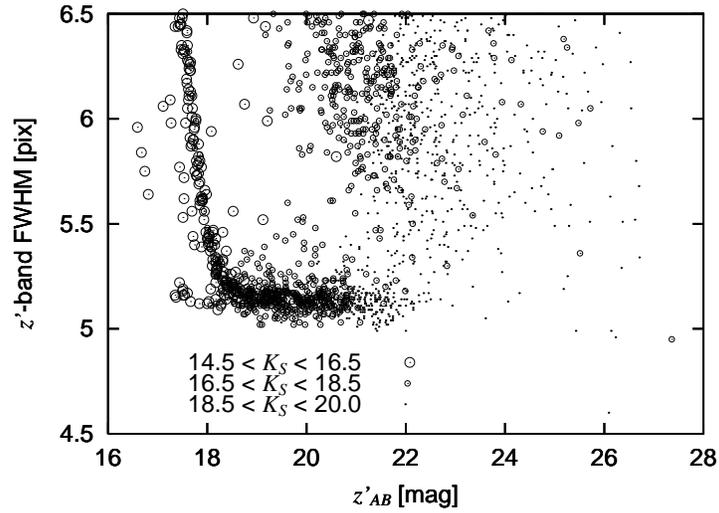}
\caption{$z'$-band FWHM vs. magnitude with the corresponding $K_S$-band objects.
Same as Figure \ref{figure:saturation_J} but for the $K_S$-band.
\label{figure:saturation_Ks}}
\end{figure}

First, we defined visible counterparts within a 2.0\arcs\ radius from the location of the $J$- or $K_S$-band selected sources.
In our results, the objects with only one optical counterpart for a $J$- and $K_S$ source are 96.8 and 98.9\%, respectively.
Objects having two or more optical counter parts are rejected from the catalog.
We also checked bright counterparts reaching to the saturation limit.
Figures \ref{figure:saturation_J} and \ref{figure:saturation_Ks} show the FWHM of the $z'$-band image as a function of $z'$-band magnitude.
Saturated point-like objects are seen at $z'<18$.
We found that optical counterparts at $J\sim17$ and $K_S\sim16.5$ reaching to the saturation limit are all point-like stars. 
Therefore we regarded objects below these saturation limits as reliable.
On the other hand, some faint objects galaxies may be identified as stars because their extended emission vanishes into the background noise.
However most objects at the faint end are visible as extended sources above the background noise (approximately more than 80\% of the total number of the objects at $J>19.5$ and $K_S>18.0$).
The classification into point-like and extended objects is therefore reasonably reliable from $J\sim17$ and $K_S\sim16.5$ to a $3\sigma$ limiting magnitude listed in Table \ref{table:obs_result}.

\begin{figure}
\epsscale{.60}
\plotone{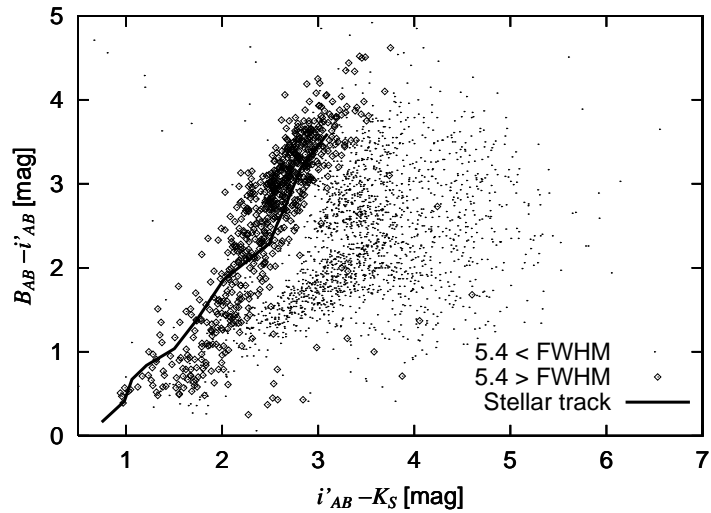}
\caption{$B-i'$ vs. $i'-K_S$ color diagram.
{The open diamonds and dots represent point-like objects of FWHM less than 5.4 pixels
and extended objects of FWHM more than 5.4 pixels, respectively.
The solid line shows the main sequence of F0-M4 stars.}
\label{figure:color_color_separation}}
\end{figure}

Secondly, we made use of the star-galaxy separation on the color-color diagram to obtain an educated guess at the object size.
Figure \ref{figure:color_color_separation} shows the $B-i'$ vs. $i'-K_S$ color diagram
which is similar to the figures of \citet{mcl95,gar95,hua97} except that we use the $i'$-band on the SDSS system \citep{fuk96}.
$B$ and $i'$ images from SUBARU/Suprime-CAM (Wada et al. 2007 in preparation) have been used to derive $B-i'$ and $i'-K_S$ colors using SExtractor AUTO magnitude.
The limiting magnitudes with a signal-to-noise ratio of five are $B\sim28.4$ and $i'\sim27.0$ in the $AB$ magnitude system, respectively.

In Figure \ref{figure:color_color_separation},
the solid line indicates the color of F0-M4 stars calculated based on the library of \citet{pic98}.
The open diamonds represent point-like objects from the catalog and flock around the stellar track.
The extended objects shown by dots usually have a redder $i'-K_S$ color.
Therefore, all objects of more than 5.4 pixels (corresponding to 1.08\arcs) in $z'$ image are regarded to be galaxies.

\begin{figure}
\epsscale{.60}
\plotone{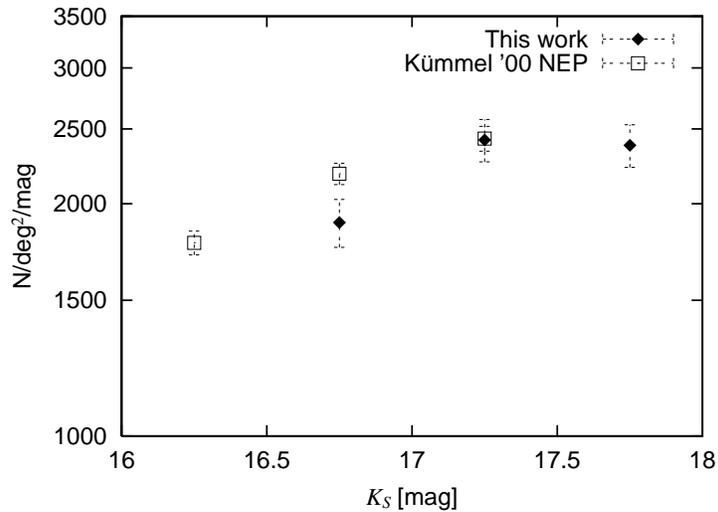}
\caption{Comparison of star counts in the NEP field.
The filled diamonds and the open squares represent the star counts of this work and \citet{kum00}, respectively.
Error bars are calculated from the number of stars in each magnitude bin using Poisson $\sqrt{N}$ counting statistics.
\label{figure:stars_comparison}}
\end{figure}

We also checked that our star-galaxy separation was reasonable by comparing other star counts in the same region.
Figure \ref{figure:stars_comparison} shows a comparison of point-like star counts.
The diamonds are our star counts and the open squares represent those of \citet{kum00}.
Our star counts at $16.5<K_S<17.0$ are lower than \citet{kum00}, 
because our star counts are classified on the basis of the optical image profiles and 
some saturated point-like objects are removed from the catalog.
However, the fainter star counts are in good agreement with each other and
stellar contamination at $K>18$ is less important \citep{kum00}.
In fact point-like objects of less than 5.4 pixels are approximately less than 20\% of the total number in our catalog at $K_S>18$ and $J>19.5$, respectively.

\subsection{Completeness correction}
\label{sub:completeness}
\begin{figure}
\epsscale{.70}
\plotone{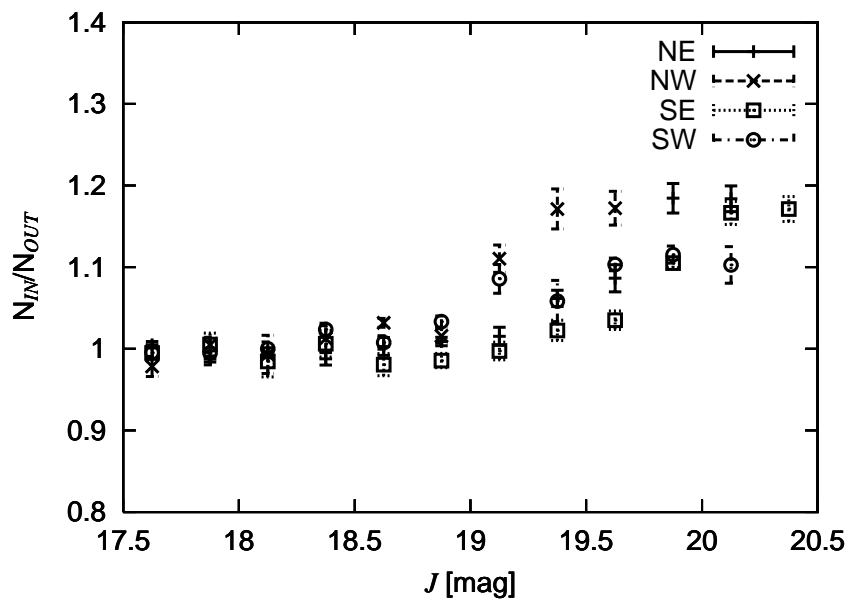}\\
\caption{The completeness correction factor, derived from Monte Carlo simulations (see text in \S \ref{sub:completeness} for details) are given for each magnitude bin.
Error bars are estimated from the standard deviation of the set of the Monte Carlo simulations.}
\label{figure:Nin_Nout_J}
\end{figure}

\begin{figure}
\epsscale{.60}
\plotone{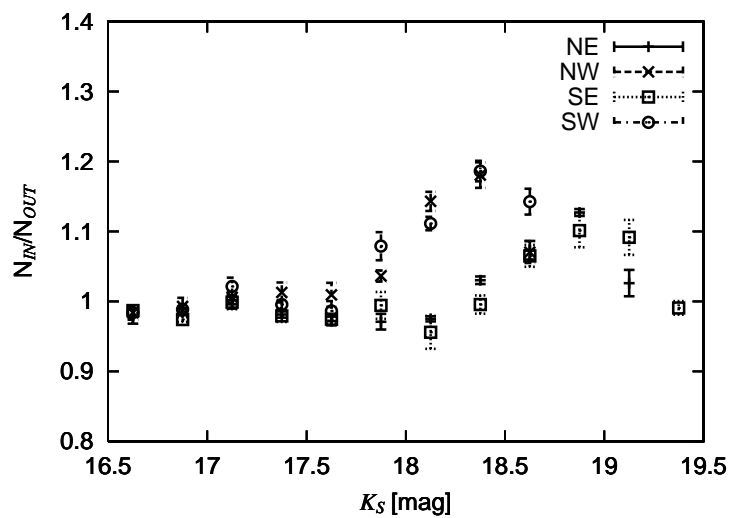}\\
\caption{The completeness correction factor for the $K_S$-band.
Symbols are the same as Figure \ref{figure:Nin_Nout_J}.}
\label{figure:Nin_Nout_Ks}
\end{figure}

\begin{figure}
\epsscale{.60}
\plotone{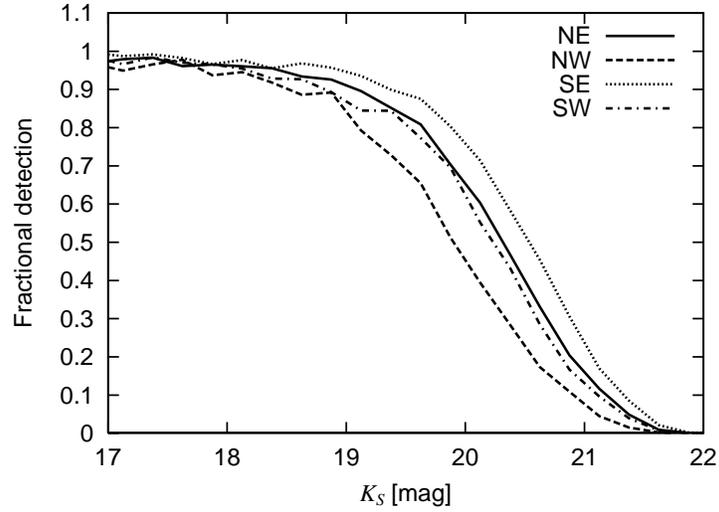}\\
\caption{The $J$-band detection completeness estimated from Monte Carlo simulations.
Each line shows the fraction of detected sources in the $J$-band for each pointing region.}
\label{figure:J-band_completeness}
\end{figure}

\begin{figure}
\epsscale{.60}
\plotone{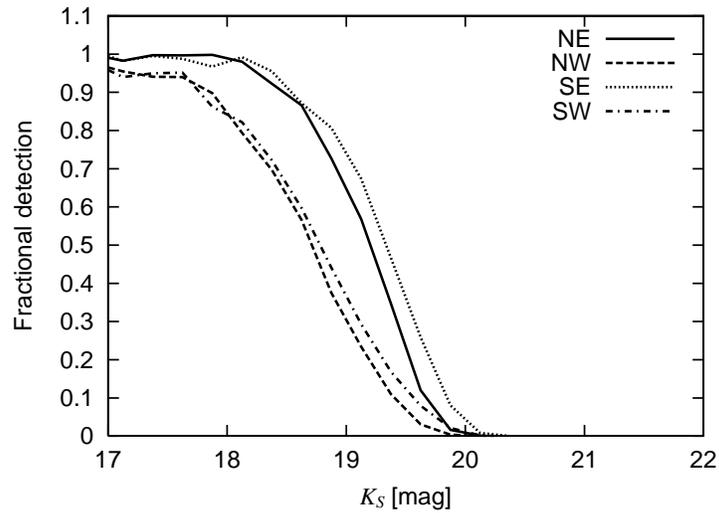}\\
\caption{The $K_S$-band detection completeness.
Meaning of the lines are the same as that in Figure \ref{figure:J-band_completeness}.}
\label{figure:Ks-band_completeness}
\end{figure}

We calculated completeness correction factors by applying Monte Carlo simulations to real images, using the STARLIST and MKOBJECTS packages within IRAF.
This approach is appropriate because we can take account of the real noise and the pixel-to-pixel sensitivity variations in the images
since randomly embedded artificial objects within the real images are extracted in the same way as that used to generate the real source catalogs.
In principle, morphological types of the galaxies influence the estimate of completeness for the faint NIR galaxies for which galaxy morphologies are unknown.
However, due to relatively large seeing size almost all the image profiles of the $J$- and $K_S$-band sources are much smaller than the seeing disk ($\sim1\arcs$, e.g. the effective radii of elliptical bulges are only $0.\arcs1-0.\arcs5$ at $z\sim1$).
Therefore, we made artificial objects (mock galaxies) as point-like.
Since too many artificial objects cause heavy blending among themselves,
we generated 25 artificial objects in the real image and
repeated this procedure 2000 times to obtain a statistically robust number of objects for each 0.25 magnitude bin.
The radii of the mock galaxies are set to the FWHM of each different pointing field listed in Table \ref{table:obs_result}.
The probability of an overlap between real and mock galaxies is less than two percent.
Objects with fluxes near the detection limit are also seriously affected by noise
and moreover, sometimes suffered from contamination from neighbor objects.
Such objects are likely to be affected by flux-boosting and may not be extracted with accurate magnitudes.
In order to judge the reliability of the source catalog, we derived the detection completeness without contamination from flux-boosting.
Figures \ref{figure:Nin_Nout_J} and \ref{figure:Nin_Nout_Ks} show the ratio of the number of embedded artificial objects in each magnitude bin to those that were actually extracted.
The NW and SW fields of the $AKARI$ NEP region are shallower than the other fields in both the $J$- and $K_S$-bands (see Table \ref{table:obs_result}) and show a 20\% difference in the completeness levels at $J\sim19.3$ and $K_S\sim18.3$, respectively (see Figures \ref{figure:Nin_Nout_J} and \ref{figure:Nin_Nout_Ks}).
However these areas are less than 30\% of the total effective area and therefore do not significantly influence the total counts.
The detection completeness as a function of magnitude are presented in Figures \ref{figure:J-band_completeness} and \ref{figure:Ks-band_completeness}.
It is noteworthy that the detection completeness is only 50-60\% at the faint end, while the inverse of completeness correction factors in Figures \ref{figure:Nin_Nout_J} and \ref{figure:Nin_Nout_Ks} are 80-100\%.
The derived completeness correction factors as well as the corrected counts for the $J$- and $K_S$-bands are listed 
in Table \ref{table:J-band_counts} and \ref{table:Ks-band_counts}, respectively.

\begin{figure}
\epsscale{.60}
\plotone{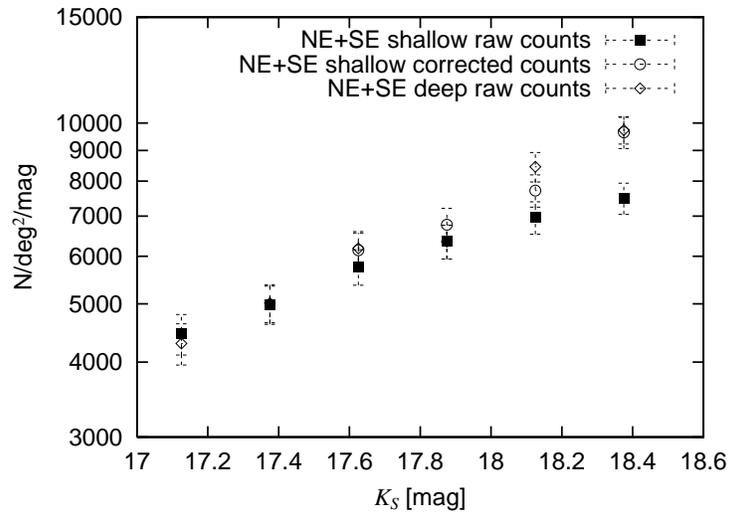}\\
\caption{Comparison of number counts to two different depths.
The open diamonds indicate the raw counts derived from all co-added frames in the NE and SE regions. 
The filled squares and the open circles show raw counts and completeness corrected counts of a subset of the NE+SE data, for which only 17\% of all frames are co-added.}
\label{figure:completeness_reliability}
\end{figure}

It is also important to evaluate the reliability of the completeness correction
because the correction factor directly affects the faint end counts.
For that purpose, we checked our completeness correction by comparing them with the completeness corrected shallow counts
which comprise of a fraction of the co-added frames and
the deep raw counts which comprise of all co-added frames \citep[a similar approach was taken by][]{vai00}.
As an example, Figure \ref{figure:completeness_reliability} shows the $K_S$-band deep raw counts in the NE+SE regions,
co-added shallow images at the same region and completeness corrected shallow counts.
The shallow images comprise of 17\% of all co-added frames correspond to $K_S\sim18.5$ at the 50\% completeness limit and are around 0.8mag shallower than those of the all co-added deep raw counts.
Note that the deep raw counts in the fainter range also require a completeness correction and
therefore cannot be directly compared with the corrected shallow counts.
From Figure \ref{figure:completeness_reliability}, it is found that the faint end of 30\% corrected shallow counts at $K_S\sim18.4$ are consistent with the deeper raw counts.

\subsection{$J$-band galaxy counts}
\label{sub:J_galaxy_counts}
In this and next subsection, we present the results for the $J$- and $K_S$-band galaxy counts in the $AKARI$ NEP field.
Since the area and depth vary among the four fields (see Table \ref{table:obs_result}),
we first obtained the galaxy number counts in each individual field
and then combined them to obtain the final number counts.
Actual numbers of galaxies and corrected counts per square degree per magnitude,
completeness correction factor and effective survey areas in square degrees in the $J$- and $K_S$-bands
are listed in Table \ref{table:J-band_counts} and Table \ref{table:Ks-band_counts}, respectively. 

In order to evaluate the slope of the $J$- band galaxy counts, we fitted power laws of the form
\begin{equation}
\label{equation:power_low}
N(mag)= a\times10^{b(mag-15)}.
\end{equation}
The best-fit slopes are then calculated in two different domains ($J>19.5$ and $J<19.5$),
since the counts in different ranges have different slopes
(the cause of the observed change in the $J$-band galaxy counts at $\sim19.5$ is discussed in \ref{sub:possible_cause}).
The derived parameters (a, b) of the fits for our and the available $J$-band galaxy counts are listed in table \ref{table:J-band_power-low}.
The slope of our galaxy counts $d(\log{N})/dm\sim0.39$ at $J>19.5$ is in good agreement with \citet{vai00}
whereas the derived slope of our galaxy counts $d(\log{N})/dm\sim0.30$ at $J<19.5$ favor the higher counts of \citet{tep99} as opposed to the lower counts of \citet{sar99}.
The $J$-band galaxy counts in the $AKARI$ NEP field show a clear change in the slope at $J\sim19.5$ from 0.39 to 0.30.

We also show no-evolution models in both the $J$- and $K_S$-bands for comparison.
The no-evolution models are pure $K$-correction models and constructed from three galaxy types (early-type, late-type 1 and late-type 2).
The $K$-corrections of each galaxy type are calculated by using the theoretical spectral energy distribution (SED) models of \citet{bru03}.
Regarding the $K_S$-band galaxy counts,
we refer to the type-dependent $K$-band luminosity function (LF) of \citet{koc01}
based on the 2MASS Second Incremental Release Catalog of Extended Sources\citep{jar00}.
However \citet{col01} point out that the $K$-band photometry of \citet{lov00} has better signal-to-noise and resolution than the 2MASS images
and enables more accurate 2MASS magnitudes to be measured than the original default magnitudes \citep[see also][]{fri05}.
We therefore normalized the characteristic magnitude $M^*_K$ of the no-evolution model at $K_S\leq13.25$ of the 2MASS counts measured by \citet{fri05} using a similar magnitude estimator to that of \citet{col01}.
On the other hand, to derive the $J$-band galaxy counts, we assume the shape of the type-dependent $J$-band LF is the same as the $K$-band LF of \citet{koc01}
and simply convert the normalized $M^*_K$ to $M^*_J$ based on the \citet{bru03} SEDs.
The basic characteristics of the SEDs i.e., galaxy type, metalicity and star formation rate are listed 
with the parameters of the LF of each galaxy type in Table \ref{table:parameters}.
The redshift of galaxy formation used for all galaxy types is $z_{form}=4$.

For the $J$-band galaxy counts,
the no-evolution model is consistent with the results of the Deep Near-Infrared Sky Survey \citep[DENIS][]{mam98}
and the bright counts of the Deep Multicolor Survey \citep[DMS][]{mar01} at magnitudes $J<17$.
However the no-evolution model is far below our observed counts from $J\sim17$ to fainter magnitudes; the factor is 2.1 at $J\sim$ 19.6.
This is because neither pure luminosity evolution (PLE) nor bulk (luminosity or number density) evolution
in the extragalactic population is included in the no-evolution model.
In particular, the inclusion of PLE into the no-evolution model will increase the normalization of the models in the intermediate magnitude range
but we reserve this work (and the investigation of bulk evolution in the near-infrared population as a whole) to the next paper in this series.

\subsection{$K_S$-band galaxy counts}
\label{sub:Ks_galaxy_counts}
\begin{figure}
\epsscale{0.75}
\plotone{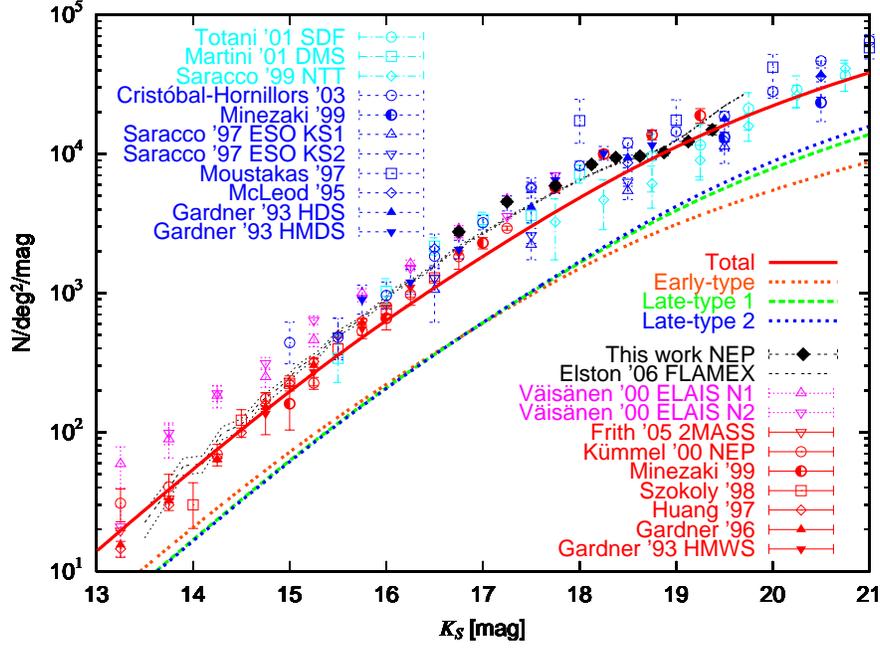}\\
\caption{$K_S$-band galaxy counts.
{The black diamonds are the galaxy counts derived from this work.
Error bars are calculated from the number of galaxies in each magnitude bin using Poisson $\sqrt{N}$ counting statistics.
Note that at the faintest magnitudes, completeness errors may also have a non-negligible effect on the source counts.
Red and magenta color symbols listed in the bottom-right corner are published counts of wide field surveys. 
The symbols in the figure correspond to,
Frith '05 2MASS \citep{fri05}, V$\rm{\ddot{a}}$is$\rm{\ddot{a}}$nen '00 ELAIS N1 and N2 \citep{vai00},
K\"{u}mmel '00 NEP \citep{kum00}, Minezaki '99 \citep{min99}, Szokoly '98 \citep{szo98},
Huang '97 \citep{hua97}, Gardner '96 \citep{gar96}, and Gardner '93 HMWS \citep{gar93}.
Blue and cyan color symbols listed in the top-left corner are from published counts of deep surveys.
The symbols correspond to
Totani '01 SDF \citep{tot01}, Martini '01 DMS \citep{mar01},
Saracco '99 NTT \citep{sar99}, Crist\'{o}bal-Hornillos '03 \citep{cri03},
Minezaki '99 \citep{min99},
Saracco '97 ESO KS1 and KS2 \citep{sar97}, Moustakas '97 \citep{mou97},
McLeod '95 \citep{mcl95} and Gardner '93 HDS and HMDS \citep{gar93}.
The FLAMEX counts of \citet{els06} are shown as the bold dotted line with the upper and lower errors
shown as the corresponding upper and lower faint dotted lines, respectively.
The counts having large errors (more than 50\% in relative errors) are not shown.
The symbols and colors for each survey are the same for the $J$-band galaxy counts.
Orange, green and blue lines indicate the no-evolution predictions of each galaxy type, described in \S \ref{sub:J_galaxy_counts}.
Red line corresponds to the total galaxy counts of these predictions.}
\label{figure:Other_Ks_counts}}
\end{figure}

The $K_S$-band galaxy counts are presented in Figure \ref{figure:Other_Ks_counts},
which shows the results in the $AKARI$ NEP field along with a compilation of published galaxy counts.
In the bright magnitude range $K_S<17.0$, our result shows slightly higher counts than \citet{gar93, kum00, els06}; by a factor of $<1.5$ at $K_S\sim$ 16.7.
The counts are however consistent with those of \citet{mcl95, mar01, vai00, cri03}.
On the other hand, though our faintest counts are approximately 50\% complete and therefore subject to large corrections,
they are in good agreement with the results of other $K$-band deep surveys \citep[except][]{sar99}.
The $K_S$-band galaxy counts in the $AKARI$ NEP field show an indication of a possible `bump' in the magnitude range $18.0<K_S<19.5$.
We discuss uncertainties and a possible cause of the observed change of the slope in the $J$- and $K_S$-band counts in \S \ref{sec:discussion}.

For the $K_S$-band galaxy counts,
although the no-evolution model is consistent with the counts derived from the wide surveys for $K_S<17$,
they are lower (approximately 35\%) than the present work at around $K_S\sim18$.
In addition, the no-evolution model cannot reproduce the characteristic change in the slope since the count slope of each galaxy component
in the no-evolution model varies rather smoothly from the bright to the faint end.

\section{$B-K_S$ Galaxy Colors}
\label{sec:galaxy_color}
\subsection{$B-K_S$ galaxy color trend}
\label{sub:color_trend}
\begin{figure}
\epsscale{.60}
\plotone{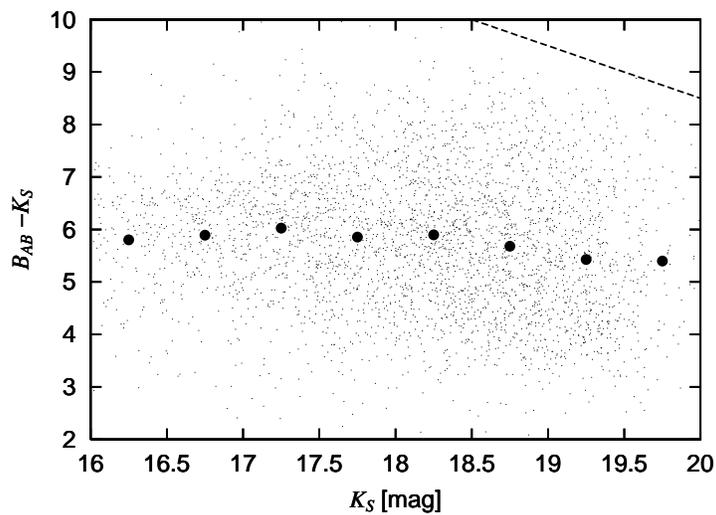}\\
\caption{$B-K_S$ galaxy color trend for the $AKARI$ NEP survey field.
Objects detected in both the $B$- and $K_S$-bands are plotted by dots.
The filled circles show the median color in each magnitude bin.
The dash line represents the region of incompleteness;
galaxies above this line are too faint to be detected in the $B$-band ($B\geq28.5$mag in AB).
\label{figure:B-Ks_color_trend}}
\end{figure}

Figure \ref{figure:B-Ks_color_trend} is a plot of the $B-K_S$ color trend of the galaxies in the $AKARI$ NEP field.
The filled circles show the median color in each magnitude bin.
All of the galaxies are detected at the 3$\sigma$ limiting magnitude in the $K_S$-band listed in Table \ref{table:obs_result}.
The median $B-K_S$ color of the galaxies becomes slightly redder up to $K_S\sim17.5$
where it reaches a maximum value of $B-K_S\sim6$,
and then tends to be bluer to fainter magnitudes.
This trend is similar to previous results \citep{gar93, mcc00, hua01}.

\subsection{$B-K_S$ galaxy color distributions}
\label{sub:color_distribution}
\begin{figure}
\epsscale{.60}
\plotone{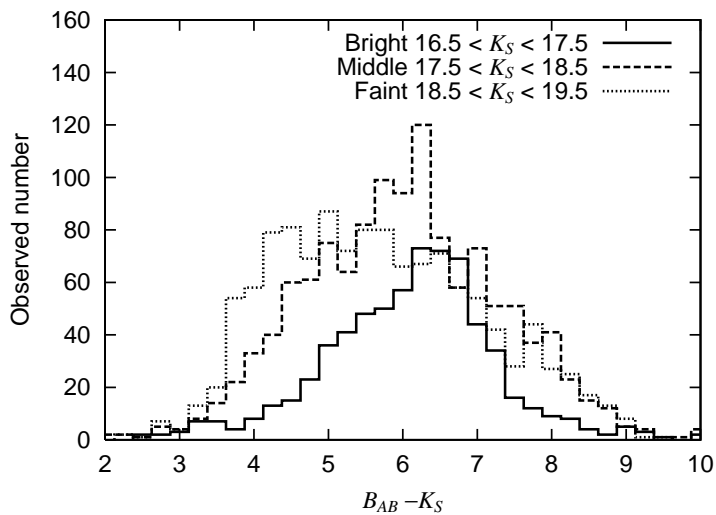}\\
\caption{$B-K_S$ galaxy color distributions for the $AKARI$ NEP survey field.
All objects are detected in both the $B$- and $K_S$-bands.
Each line corresponds to selected objects in a different magnitude bin.
\label{figure:B-Ks_color_distribution}}
\end{figure}

Figure \ref{figure:B-Ks_color_distribution} shows the $B-K_S$ galaxy color distributions.
The solid line is the bright galaxy sample selected in $16.5<K_S<17.5$,
dashed and dotted lines correspond to the galaxy samples selected in $17.5<K_S<18.5$ and $18.5<K_S<19.5$, respectively.
These samples cover the bump region of the $K_S$-band galaxy counts described in \S\ref{sub:Ks_galaxy_counts}.
The number of sources in the bright sample ($16.5<K_S<17.5$) increases from $B-K_S\sim4$ to a peak at $B-K_S\sim6.5$ and decreases thereafter.
The decrease of the red end is steeper than the increase at the blue end.
Though the peak of middle sample ($17.5<K_S<18.5$) is somewhat shifted toward a bluer $B-K_S$ color of 6.2 and
the number of blue and red galaxies are relatively larger than the bright sample,
the shape of the $B-K_S$ color distribution of the middle sample is similar to that of the bright sample.
On the contrary, the shape of the faint sample ($18.5<K_S<19.5$) is obviously different from the others, with a dramatic increase in the number of blue galaxies at $3.5<B-K_S<5.0$.
We confirm this difference in the two samples by applying a $Kolmogorov$-$Smirnov$ test ($K$-$S$ test) to our data.
The $K$-$S$ test results in a probability of $10^{-18}$ that two (bright and faint) galaxy samples are identical.

\section{Discussion}
\label{sec:discussion}
We presented the change of the $J$- and $K_S$-band galaxy count slopes and the bluer trend in the $B-K_S$ color distribution
in \S \ref{sub:J_galaxy_counts}, \S \ref{sub:Ks_galaxy_counts} and \S \ref{sec:galaxy_color}, respectively.
In this section, we discuss the relation ship of $J$- and $K_S$-band count slopes (\ref{sub:relation}),
the source of the uncertainties in the galaxy counts (\ref{sub:uncertainty})
and implications from other observational studies (\ref{sub:possible_cause}).

\subsection{Relation of $J$- and $K_S$-band galaxy counts slopes}
\label{sub:relation}
In the $J$-band, we presented galaxy counts based on the largest sample to date which bridges the previous deep and shallow galaxy survey counts.
We showed the clear change of the slope at $J\sim19.5$ from $d(\log N)/dm=0.39$ to 0.30.
However the change in the slope of the $J$-band counts is less prominent than that in the $K_S$-band counts where it is visibly seen as a `bump'.

The brighter component of the $K_S$-band counts (above the `bump') is dominated by early-type galaxies obeying the PLE \citep[e.g.][]{met91,met96,poz96}.
The less prominent slope change in the $J$-band counts is due to the early-type galaxies having large $J-K_S$ colors (i.e. fainter $J$-band magnitudes).
Indeed, \citet{sar99} reported $J-K>1.5$ at $K_S<18$, while \citet{ber98} discussed that early-type population show $J-K>2$ at $z>1$. A $J-K$ color of $\sim$1.5 would imply to a corresponding turnover in the $J$-band counts at $J\sim$19.5.

\subsection{Uncertainties in the $K_S$-band count}
\label{sub:uncertainty}
For the $K_S$-band galaxy counts in the $AKARI$ NEP field, we have shown an indication of a change in the slope over the magnitude range $18<K_S<19.5$.
Here we discuss the uncertainties in the $K_S$-band counts over three different magnitude ranges: $18>K_S$, $18<K_S<19$, $19<K_S$,
and compare in particular with the FLAMEX counts \citep{els06}, a similar NIR survey covering 7.1 deg$^2$ using the same instrument (FLAMINGOS).

In the bright magnitude range ($K_S<18.0$),
the slope of fitted power law of equation \ref{equation:power_low} is $d(\log{N})/dm=0.32\pm0.06$.
Our result shows higher counts than the results of FLAMEX.
The difference between the two surveys is a factor 1.3 at $K_S\sim$ 16.7.
This may be attributed to the fact that the two surveys adopted a different approach for the star-galaxy separation.
Star-galaxy separation of FLAMEX was performed by using color-color diagrams
while in the case of this work, the separation is based on the image profiles of optical counterparts.
Therefore some stars with profiles of more than 5.4 pixels may be included in the galaxy counts.
However star counts in the NEP field do not heavily affect the galaxy counts in the fainter magnitude range, $K>18$ \citep{kum00}.
In fact point-like objects of less than 5.4 pixels are less than 20\% of the total number in our catalog at $K_S>18$.
Thus the observed turn over from $K_S\sim17$ is considered an intrinsic feature.

In the intermediate magnitude range ($18.0<K_S<19.0$),
the slope of our galaxy counts is $d(\log{N})/dm=0.11\pm0.02$.
Our $K_S$-band counts show a flattening and are in good agreement with the FLAMEX results.
The tendency of the flattening is also seen in the results of FLAMEX and other deep surveys \citep[e.g.][]{roc03, cap05, for06}.

In the faint magnitude range ($K_S>19.0$), 
the slope of our galaxy counts is $d(\log{N})/dm=0.32\pm0.06$.
Our faintest counts may be significantly affected by incompleteness (approximately 50\% complete) and show lower counts than the results of FLAMEX; the factor is 1.3 at $K_S\sim$ 19.3.
However, they are consistent with the results of other $K$-band deep surveys \citep{gar93, mcl95, tot01, cri03}.
We therefore cannot rule out the possibility that some of the reason for the change of slope observed over the magnitude range $18.0<K_S<19.5$
may be due to incompleteness since the completeness corrections are non-negligible at these magnitudes.

\subsection{Evidence for number density evolution in the early-type population}
\label{sub:possible_cause}
To reproduce the turn over in the slope at $K_S\sim17.5$, \citet{cri03} delayed the galaxy formation epoch for elliptical and spiral galaxies to quite low redshift, $z_{form}<2$
and demanded the presence of a population of star-forming dwarfs at all redshifts in a $\Lambda$-dominated universe. 
However, since numerous early-type galaxies at $z>2$ have recently been found \citep[e.g.][]{dad05,lab05},
it is unrealistic to completely extinguish early-type population at $z>2$.

Our result of the $B-K_S$ color distribution also shows the number of blue galaxies dramatically increases below the `bump' to fainter $K_S$ magnitudes.
These results indicate that the contribution of early-type galaxies to the counts is relatively low in the fainter magnitude range,
a result that has been suggested by several studies including galaxy morphologies at $z>1$.
\citet{kaj01} reported that the comoving density of $M_V<-20$ galaxies decreases at $z>1$ in each morphological category based on the result of the HST WFPC2/NICMOS archival data of the HDF-N.
In particular, they showed that the number density of the bulge-dominated galaxies conspicuously decreases to 1/5 between the $0.5<z<1.0$ and $1.0<z<1.5$ bins.
Other morphological studies also show the lack of massive ellipticals at redshifts greater than $z\sim1.5$ \citep[e.g.][]{fra98, rod01}.

Moreover, a similar decrease in the number of red galaxies at higher redshift is indicated from studies of extremely red objects (EROs).
The EROs are selected to have very red optical-to-near infrared colors such as $R-K>5-7$, $I-K>4-6$.
The number counts of EROs using a sample ($K<21$) of 158 EROs in the ELAIS N2 field shows lower counts
than those predicted by models in which all E/S0 galaxies evolve according to PLE. However, the counts can be fit by a model combining PLE with galaxy merging and a decrease in the comoving number density of these passive galaxies with redshift \citep{roc02}.
Based on a very deep sample ($K_S<22$) of 198 EROs from the ESO/GOODS data in the CDF-S,
\citet{roc03} reported that the ERO number count flattens at $K>19.5$ from $d(\log N)/dm=0.59$ to 0.16.

This negative density evolution also has been reported from studies of the evolution of the $K$-band LF at various redshifts: $z<1$
\citep{gla95, feu03}, $z<2$ \citep{bol02, poz03}, $z<3.5$ \citep{kas03} and $z<4$ \citep{sar06}.
Note that \citet{poz03} show a contrary result, however, their large errors due to the limited sample size do not rule out a decrease of the comoving number density with redshift.

To summarize, the number density evolution of the elliptical galaxy population remains unsolved.
However the bump in the $K_S$-band galaxy count is an important clue and imposes tight constraints on the formation and evolution of each galaxy population
because galaxy counts are measured with less indefiniteness than other observables and wider and deeper NIR survey can pin down the slope of the galaxy counts more precisely.
We entrust further discussion about the evolution of each galaxy population to our forthcoming paper.

\section{Summary}
\label{sec:sammary}
We performed a NIR survey in the $AKARI$ NEP field, using FLAMINGOS on the KPNO 2.1m telescope.
The total area is 750 arcmin$^2$ and the limiting magnitudes to a signal-to-noise ratio of three in the deepest regions are 21.85 and 20.15 in the $J$- and $K_S$-bands respectively in the $Vega$ magnitude system.
The extracted objects are classified into stars and galaxies by using image profiles of their optical counterparts from Subaru Suprime-cam observations.
The resultant star counts are consistent with other published results in the same region.
Completeness corrections are estimated by adding artificial sources to the images and extracting them.
The reliability of the completeness corrections are also evaluated by comparing the completeness corrected shallow counts and the deep raw counts.
We then presented the $J$- and $K_S$-band galaxy counts and the $B-K_S$ color distribution in three different magnitude slices down to $K_S=19.5$.
Our $J$- and $K_S$-band galaxy counts are broadly in agreement with those of other published results, while our counts indicate a change in the slope in both the $J$- and $K_S$-band counts.
The slope of our $J$-band galaxy counts changes at $J\sim19.5$ from $d(\log{N})/dm=0.39\pm0.02$ to $0.30\pm0.03$, while the $K_S$-band galaxy counts show a more significant `bump' in the magnitude range $18.0<K_S<19.5$.
The $B-K_S$ color distribution of our galaxy samples also shows an obvious change around the `bump' magnitude of the $K_S$-band counts.
The number of blue galaxies in the magnitude range $18.5<K_S<19.5$ is significantly larger than that of the $K_S<17.5$ samples.

The authors wish to thank the referee who's comments greatly improved the contents of this work.

\clearpage

\clearpage

\begin{figure}
\epsscale{0.75}
\plotone{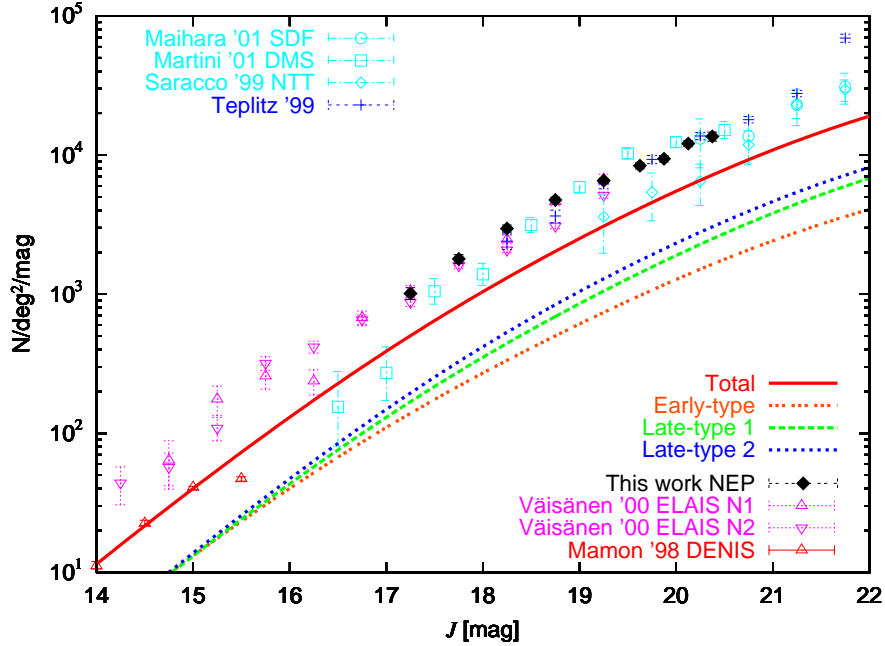}\\
\caption{$J$-band galaxy counts.
{The black diamonds are galaxy counts derived from this work.
Error bars are calculated from the number of galaxies in each magnitude bin using Poisson $\sqrt{N}$ counting statistics.
Note that at the faintest magnitudes, completeness errors may also have a non-negligible effect on the source counts.
The symbols in the figure correspond to, 
Maihara '01 SDF \citep{mai01}, Martini '01 DMS \citep{mar01},
V$\rm{\ddot{a}}$is$\rm{\ddot{a}}$nen '00 ELAIS N1 and N2 \citep{vai00}, Saracco '99 NTT \citep{sar99}, Teplitz '99 \citep{tep99} and Mamon '98 DENIS \citep{mam98}.
The counts having large errors (more than 50\% in relative errors) are not shown.
The symbols and colors for each survey are the same for the $K_S$-band galaxy counts.
Orange, green and blue lines indicate the no-evolution predictions for each galaxy type, described in \S \ref{sub:J_galaxy_counts}.
Red line corresponds to the total galaxy counts of these predictions.}
\label{figure:Other_J_counts}}
\end{figure}

\clearpage

\begin{table}
\begin{center}
\caption{Observed center coordinates and areas.}
\label{table:obs_area}
\begin{tabular}{ c c c c }
\tableline \tableline
 Region name & R.A.          & Dec.       & Area         \\
             & (J2000)       & (J2000)    & (arcmin$^2$) \\
    (1)      & (2)           & (3)        & (4) \\ \tableline
    NE       & 17h 55m 48.2s & +66d 45\arcm\ 41\arcs & 234 \\
    NW       & 17h 53m 41.3s & +66d 45\arcm\ 40\arcs & 91  \\
    SE       & 17h 55m 47.8s & +66d 30\arcm\ 41\arcs & 306 \\
    SW       & 17h 53m 42.3s & +66d 30\arcm\ 43\arcs & 119 \\ \tableline
\end{tabular}
\tablecomments{(2) Unit of R.A. is hours, minutes, and seconds.
(3) Unit of Dec. is degrees, arcminutes, and arcseconds.} \\
\end{center}
\end{table}

\begin{table}
\begin{center}
\caption{Observational conditions of NIR survey in the $AKARI$ NEP field.}
\label{table:obs_result}
\begin{tabular}{ c c c c c } 
\tableline \tableline
\multicolumn{5}{c}{$J$-band} \\ \tableline
 Field name & Exposure time & FWHM 		& 50\% detection completeness & 3$\sigma$ limit \\
			& (s)			& (arcsec)	& (magnitude)		& (magnitude) \\
 (1)   		& (2)			& (3)  		& (4)				& (5) \\ \tableline
 NE   		& 9480			& 1.38 		& 20.4				& 21.6 \\
 NW  		& 13080			& 1.86 		& 20.0				& 21.2 \\
 SE			& 8400			& 1.08 		& 20.5				& 21.9 \\
 SW			& 10920			& 1.50 		& 20.3				& 21.5 \\ \tableline
\multicolumn{5}{c}{$K_S$-band} \\ \tableline
 NE  		& 6160			& 1.08		& 19.2				& 19.9 \\
 NW  		& 6950			& 1.80 		& 18.8				& 19.4 \\ 
 SE  		& 7110			& 1.08 		& 19.4				& 20.2 \\ 
 SW  		& 7980			& 2.04 		& 18.8				& 19.5 \\ \tableline
\end{tabular}
\tablecomments{(2) Total exposure time for each pointing field. 
(4) Measurement of completeness is described in \S\ref{sub:completeness}.
(5) Aperture radius is set a threes of half of FWHM.} \\
\end{center}
\end{table}

\begin{table}
\begin{center}
\caption{Description of the SEDs, luminosity function parameters and rest-frame colors used in the models.}
\label{table:parameters}
\begin{tabular}{ c c c c c c c}
\tableline \tableline
 Galaxy type 	& Metalicity	& Star formation rate	& $\phi^*$					& $\alpha$	& ($J-K_S$) \\
				&(Z/Z$_{\odot}$)& (Gyr)					& (Mpc$^{-3}\times10^{-3}$)	&			&			\\
    (1)     	& (2)          	& (3)   	     		& (4)						& (5)		& (6)		\\ \tableline
 Early-type		& 2.5			& Exp. $\tau=0.5$		&  1.3						& -0.8		& 1.0		\\
 Late-type 1	& 1.0			& Exp. $\tau=1.0$		&  2.0						& -0.9		& 0.9		\\
 Late-type 2	& 0.4			& Exp. $\tau=2.0$		&  2.0						& -0.9		& 0.9		\\ \tableline
\end{tabular}
\tablecomments{(3) The $e$-folding time of the star formation rate.
(4) The parameter $\phi^*$ is a normalization coefficient that has the dimensions of the number density of galaxies.} \\
\end{center}
\end{table}

\begin{table}
\begin{center}
\caption{$J$-band raw and corrected galaxy counts.}
\label{table:J-band_counts}
\begin{tabular}{ c c c c c c }
\tableline \tableline
 $J$ mag	& N$_{raw}$	& Completeness	& N$_{corr}$	& Error	& Area  \\
 (magnitude)& (mag$^{-1}$deg$^{-2}$) & 	& (mag$^{-1}$deg$^{-2}$)& (mag$^{-1}$deg$^{-2}$) & (arcmin$^2$) \\
 (1)        & (2)		&  (3) 	& (4)   & (5)   & (6) \\ \tableline
 17.250		& 989		& 0.98	& 1000	& 100	& 750 \\ 
 17.750		& 1728		& 0.96	& 1800	& 130	& 750 \\ 
 18.250		& 2822		& 0.95	& 2970	& 170	& 750 \\ 
 18.750		& 4454		& 0.94	& 4700	& 200	& 750 \\ 
 19.250		& 5961		& 0.91	& 6600	& 300	& 750 \\ 
 19.625		& 7238		& 0.87	& 8300	& 500	& 750 \\ 
 19.875		& 7910		& 0.84	& 9400	& 500	& 659 \\ 
 20.125		& 9046		& 0.75	& 12100	& 600	& 659 \\ 
 20.375		& 8988		& 0.66	& 13600	& 1100	& 306 \\ \tableline
\end{tabular}
\tablecomments{
(1) J-band magnitude at each magnitude bin.
(2) Raw galaxy counts detected in this magnitude range.
(4) Completeness correction factors.
(3) Completeness corrected counts.
(5) Errors are dominated by Poisson $\sqrt{N}$ statistical errors.
Errors in the completeness correction factor itself are found to be negligible.
(6) Total effective area.}
\end{center}
\end{table}

\begin{table}
\begin{center}
\caption{Comparison of the power-law slopes of the $J$-band galaxy counts.}
\label{table:J-band_power-low}
\begin{tabular}{ c c c c c c }
\tableline \tableline
 $J<19.5$					& a				& b				& Range			& Area \\
							&				&				& (magnitude)	& (arcdeg$^2$) \\
 (1)						& (2)			& (3)			& (4)			& (5) \\ \tableline
 NIR $AKARI$ NEP survey		& 150$\pm$30	& 0.39$\pm$0.02	& 17.0-19.5	& $2.1\times10^{-1}$ \\
 \citet[][ELAIS N1]{vai00}	& 140$\pm$30	& 0.40$\pm$0.03	& 17.0-19.5	& 2.0-$0.6\times10^{-1}$ \\
 \citet[][ELAIS N2]{vai00}	& 130$\pm$30	& 0.38$\pm$0.02	& 17.0-19.5	& 5.0-$2.9\times10^{-1}$ \\ \tableline
 $J>19.5$					&				&				&				& \\ \tableline
 NIR $AKARI$ NEP survey		& 330$\pm$140	& 0.30$\pm$0.03 & 19.5-20.5 	& 2.1-$0.9\times10^{-1}$ \\
 \citet{tep99}				& 430$\pm$150	& 0.28$\pm$0.03 & 19.5-21.0 	& $5.3\times10^{-2}$ \\
 \citet[][ESO NTT]{sar99}	& 80$\pm$120	& 0.37$\pm$0.11 & 19.5-21.0		& $5.5\times10^{-3}$ \\ \tableline
\end{tabular}
\tablecomments{(2), (3) The ampliltude of the fitting power law given by eq. \ref{equation:power_low}.
(4) Magnitude range of the fitted data.
(5) Effective area of each survey.}
\end{center}
\end{table}

\begin{table}
\begin{center}
\caption{$K_S$-band raw and corrected galaxy counts.}
\label{table:Ks-band_counts}
\begin{tabular}{ c c c c c c }
\tableline \tableline
 $K_S$ mag	& N$_{raw}$	& Completeness	&N$_{corr}$	& Error	& Area  \\
 (magnitude)& (mag$^{-1}$deg$^{-2}$)	&	& (mag$^{-1}$deg$^{-2}$)& (mag$^{-1}$deg$^{-2}$) & (arcmin$^2$) \\
 (1)        & (2)		& (3)   & (4)	& (5)   & (6) \\ \tableline
 16.750		& 2707		& 0.98	& 2760	& 160	& 750 \\ 
 17.250		& 4406		& 0.97	& 4500	& 200	& 750 \\ 
 17.750		& 5702		& 0.96	& 5900	& 200	& 750 \\ 
 18.125		& 8025		& 0.96	& 8400	& 400	& 750 \\ 
 18.375		& 8908		& 0.94	& 9400	& 400	& 750 \\ 
 18.625		& 8390		& 0.87	& 9600	& 500	& 750 \\ 
 18.875		& 9067		& 0.89	& 10200	& 600	& 540 \\ 
 19.125		& 8427		& 0.69	& 12200	& 700	& 540 \\ 
 19.375		& 7435		& 0.50	& 15000	& 1200	& 306 \\ \tableline
\end{tabular}
\tablecomments{
Description of the counts is given in the footnote of Table \ref{table:J-band_counts}}
\end{center}
\end{table}

\end{document}